\documentstyle[12pt]{article}
\begin{document}
\setlength{\baselineskip}{.7cm}
\renewcommand{\thefootnote}{\fnsymbol{footnote}}
\sloppy

\begin{center}
{\Large \bf Mapping Self-Organized Criticality onto Criticality}

\vskip .17in
Didier Sornette, Anders Johansen and Ivan Dornic

{\it Laboratoire de Physique de la Mati\`ere Condens\'ee, CNRS
URA190\\
 Universit\'e des Sciences, B.P. 70, Parc Valrose, 06108 Nice Cedex 2, France }

\vskip .2in

\end{center}
\bigskip


        05.70.Ln : Nonequilibrium thermodynamics, irreversible processes

        05.40+j  :  Fluctuations phenomena, random processes and brownian
motion

\vskip 1cm

{\bf Abstract :}
We present a general conceptual framework for self-organized criticality
(SOC), based on the recognition that it is nothing but the expression,
''unfolded'' in a suitable parameter space, of an underlying {\it
unstable} dynamical critical point. More precisely, SOC is shown to
result from the tuning of the {\em order parameter} to a vanishingly
small, but {\em positive} value, thus ensuring that the corresponding
control parameter lies exactly at its critical value for the underlying
transition. This clarifies the role and nature of the {\em very slow
driving rate} common to all systems exhibiting SOC. This mechanism is
shown to apply to models of sandpiles, earthquakes, depinning, fractal
growth and forest-fires, which have been proposed as examples of SOC.

\vskip 1cm

{\bf R\'esum\'e :}
Nous proposons une strat\'egie g\'en\'erale pour identifier le m\'ecanisme
responsable
des ph\'enom\`enes critiques auto-organis\'es, bas\'ee sur l'id\'ee qu'ils sont
simplement la traduction, dans un espace de param\`etres choisis, d'un
point critique dynamique instable standart.
La criticalit\'e auto-organis\'e r\'esulte
du contr\^ole du param\`etre d'ordre ajust\'e \`a une valeur
positive tendant vers z\'ero, ce qui assure automatiquement que le
param\`etre de contr\^ole
correspondant se cale exactement sur sa valeur critique de la transition de
critique sous-jacente. Ce r\'esultat explique le r\^ole particulier
jou\'e par le
for\c cage infiniment lent qui est un caract\`ere commun \`a tous les
syst\`emes
critiques auto-organis\'es. Nous appliquons ces id\'ees aux mod\`eles de
tas de sable,
aux mod\`eles de tremblements de terre, de feux de for\^{e}ts, aux
transitions de d\'ecrochage et aux mod\`eles
de croissance fractale, qui ont \'et\'e propos\'es comme autant d'exemples
caract\'eristiques de la criticalit\'e auto-organis\'ee.

\pagebreak

\section{Introduction}

Following a number of early insightful studies [1-6], the past decade
has witnessed a clear acknowledgement that
many natural phenomena must be described by power law statistics.
Correspondingly, an intense activity has developed in order to
understand the origin of these ubiquitous power law tails. This has led
in particular to the concept of `self-organized criticality' (SOC)
[7-8], according to which certain dynamically driven spatially extended
systems evolve spontaneously towards a {\it critical} globally
stationary dynamical state with no characteristic time or length scales.

The fundamental idea underlying SOC is that, unlike phase transitions in
equilibrium statistical physics, the critical state is reached without the
need of fine-tuning a control parameter, i.e. the critical state is an
attractor of the dynamics. To illustrate the basic ideas of SOC, Bak and
co-workers used a cellular automaton inspired from the creation of
avalanches in a pile of sand. In these models, "sand" are added grain by
grain on a lattice until a local slope becomes unstable, and an avalanche
is initiated. In this way, the pile reaches a stationary critical state,
characterized by a critical slope, in which additional grains of sand
will fall off the pile via avalanches of all sizes from grain to lattice
scale, distributed in size and lifetime according to a power law.

In spite of a large theoretical effort, the general conditions under
which a physical system exhibits SOC are still largely unknown. Some
facts have however been established :

\begin{itemize}
\item some systems qualify as SOC if their large scale
evolution obeys a diffusion equation (possibly non-linear but with no
characteristic time
scale) which satisfies a global conservation law [9,10] ;

\item There exists a class of systems which exhibit diffusion-like response
but do not
obey a global conservation law and nevertheless seem to exhibit SOC
[11-14]. In these
cases, the underlying mechanism for SOC is still not clear even if there
seems to exist a
deep relationship with the problem of synchronization of coupled threshold
oscillators of relaxation occurring in some domain of the parameter space
[8,15-18].

\item More generally, a feedback mechanism must operate which, from the
perspective
of usual critical phenomena, describes the action of the order parameter
onto the control
parameter [19] and attracts the dynamics to a
critical state. This then suggests a mechanism for transforming
usual ''unstable critical
phase transitions'' into SOC [19]. We call unstable those critical phenomena
which are not self-organized.

\end{itemize}

In summary, the present theoretical understanding of SOC is rather
fragmented with no real unifying perspective.
Our goal here is to attempt to present a general theoretical framework for
SOC, based on
the recognition that it is nothing but the expression, ''unfolded'' in a
suitable
parameter space, of an underlying genuine {\it unstable} critical point. This
correspondence provides, what we believe is, the fundamental mechanism for
SOC and more
information on the relevant critical exponents can be obtained within this
framework.

\section{The Nature of Self-Organized Criticality}

\subsection{General mechanism}

The essence of our approach can be summarized in a few sentences.
Consider a ''standard'' {\it unstable} critical phase transition, such
as the Ising ferromagnet or, analogously, bond percolation. Here, a
spin, up or down, is assigned to each site with an exchange coupling
constant $J$. Furthermore, one defines two nearest neighbour sites to be
connected with a probability $p=1-e^{-2J/K_BT}$ if both have spin up.
For zero external field, this defines a critical temperature $T_c$ or
bond-density $\rho_c$ below which the order parameter $m_0$, the
magnetization or the probability of an infinite cluster, is zero and
behaves as $m_0 \propto (T_c-T)^{\beta}$ above.
The transition is further characterized by a
diverging correlation length $\xi \propto (T-T_c)^{-\nu}$ and
susceptibility $\chi \propto (T-T_c)^{-\gamma}$ as $T_c$ is approached,
hence quantifying the spatial fluctuations of the order parameter.
Suppose now that it turns out to be natural for the system under
consideration that, instead of controlling $T$, the ''operator''
controls the order parameter $m_0$ and furthermore takes the limiting
case of fixing it to a positive but arbitrary small value. The condition
$m_0 \rightarrow 0^+$ is equivalent to $T \rightarrow T_c^-$.
(Specifically, the scenario above comes more natural
in the context of a "zero strength" infinite cluster, where only one
bond needs to be broken, than that of a ferromagnet.) In other words,
the system is at the critical value of the {\it unstable} critical point
and must therefore exhibit fluctuations at all scales in its response.
This is nothing but the hallmark of the underlying {\it unstable}
critical point. As will be shown explicitly in the following examples,
this scenario applies most naturally to out-of-equilibrium driven
systems.

More precisely, we shall argue that systems exhibiting SOC present a
genuine critical transition when forced by a suitable control
parameter, often in the form of a generalized force (torque for sandpile
models, stress for earthquake models, force for depinning systems).
Then, SOC appears as soon as one controls or drives the system via the
order parameter of the critical transition (it turns
out that in these systems, the order parameter is also the conjugate of
the control parameter in the sense of mechanical Hamilton-Jacobi
equations). The order parameter hence takes in general the form of a
velocity or flux. The condition that the driving is at $M \rightarrow
0^+$ illuminates the special role played by the constraint of a {\it
very slow driving rate} and {\ common to all systems exhibiting SOC, as
being the exact condition to control the order parameter at $0^+$ which
ensures the positioning at the exact critical value of the control
parameter.

We shall now illustrate and develop this general idea, by a detailed
discussion of four examples, the sandpiles, earthquake models,
pinned-depinned lines or Charge-Density-Waves models, fractal growth
processes and forest-fires.

\subsection{The sandpile}

\subsubsection{Model of an unstable transition}

Let us consider the cellular automaton sandpile model [7] and put it in
a geometry inspired from experiments [20], namely that of a rotating
cylinder. The cylinder axis is horizontal and is the same as the rotation
axis. The
cylinder is partially filled  with "sand" presenting an initially flat
horizontal interface below the axis. Suppose that the axis of the cylinder
is held to
a fixed frame by a torsion spring on which one can exert a controlled
torque $T$. If
$T=0$, the cylinder takes the position such that the surface of the sand is
horizontal, {\it i.e.}, the rotation angle $\theta =0$. If one starts to
exert a
non-vanishing $T$, the cylinder rotates up to an angle $\theta$ such that
the torque
exerted by the tilted sandpile balances exactly the applied $T$. Increasing
$T$, one
finally reaches a critical value $T_c$ at which the sandpile reaches its slope
$\theta_c$ of instability corresponding to the triggering of sand flow $J$,
whose
magnitude increases with $T>T_c$.

We witness a critical {\it sliding} transition, from a state of repose
($J=0$ for $T<T_c$) to an active sliding state ($J>0$ for $T>T_c$)
corresponding to an average steady rotation of the cylinder at a
non-zero average angular velocity $<\frac{d\theta}{dt}>$. This rotation
occurs, since increasing $T$, the slope and therefore torque exerted by
the sand can no more increase and balance the applied torque. One thus
enters a dynamical regime in which the avalanches overlap and construct
a non-zero fluctuating flow.

This critical transition is characterized by the way the average flow increases
from zero above $T_c$ ($J \sim (T-T_c)^{\beta}$), as well as by the spatial
and temporal
correlations in the local burst of non-zero $J$ below $T_c$. The maximum
size of these
bursts when increasing $T$ by a small amount below $T_c$ (or by inserting a
small
perturbation such as a local grain-hole addition) allows one to define the
correlation
length $\xi^-$. Note that, from the constraint of grain conservation, the
sand flux $J$
and the cylinder angular velocity $<\frac{d\theta}{dt}>$ are simply
proportional and
describe physically the same ordered response of the system to the
increasing control
parameter $T$  ($J \sim <\frac{d\theta}{dt}> = 0$ for $T<T_c$ and $J
\sim <\frac{d\theta}{dt}> \  > 0$ for $T>T_c$). For our purpose, it is now more
illuminating to speak of the order parameter in terms of
$<\frac{d\theta}{dt}>$.

The above critical {\it sliding} transition occurs when controlling the
torque $T$.
However, suppose that one controls instead the angular velocity
$\frac{d\theta}{dt}$ and
impose $\frac{d\theta}{dt}=0^+$, i.e. a vanishingly small positive value. In
the
language of mechanics, this corresponds to interchange the role of the
conjugate
variables $T$ and $\frac{d\theta}{dt}$ (defined by the fact that their
product gives the
mechanical power of the system). It is then clear that $T$ adjusts to its
critical value
$T_c$ and the
response of the sandpile will just be that documented previously [7],
in terms of power law distribution of avalanches.

Let us briefly conclude this section of the sandpile paradigm by
describing how a specific model for the critical sliding instability can
be implemented, knowing the initial rules of the sandpile cellular
automaton. A given sand configuration is characterized by the set of
column heights and slopes, the total torque just being the sum over all
sites of the torque exerted by each column with respect to the cylinder
axis of rotation. Then, a given grain configuration corresponds to a
global torque $T$ (it is clear that many configuration have the same $T$
similarly to the finding that many micro-states correspond to the SOC
macro-state, as found by [22] in a study of Abelian sandpile cellular
automata). Then, a small increment of $T$ corresponds to a global
increase of the local slopes possibly leading to local instabilities,
{\it i.e.}, avalanches. Starting from a small initial $T$, the avalanche
regime is only transient. When one reaches and overpasses $T_c$, the
continuous flow appears and the order parameter ("sand" flow) becomes
positive, the avalanches being order parameter fluctuations. Thus, the
size of the characteristic avalanches is the correlation length. Note
that in contrast to a previous discussion of the connection between SOC
and standard critical phenomena [21], our framework introduces naturally
the current as the order parameter. The slope is just a dynamical
variable which adjusts itself as a function of the external torque
(control parameter).

\subsubsection{Scaling laws}

Viewed as a critical transition, one can define the corresponding power
laws :
$$
J \sim (T-T_c)^{\beta}   \eqno{(1)}
$$
for the current flow (order parameter),
$$
\chi \sim |T_c-T|^{-\gamma}   \eqno{(2)}
$$
for the susceptibility or the response function to (i.e. current flow
induced by) a small perturbation very close to the critical transition
[21] and
$$
\xi  \sim |T_c-T|^{-\nu}   \eqno{(3)}
$$
for the correlation length given by the linear size of the domain which
is sensitive to a local perturbation. Note that we assume here that
$\gamma$ and $\nu$ are the same on both sides of $T_c$.

In general, one should expect these three exponents to be interdependent
and obey a scaling relation expressing that the local flux induced by a
perturbation is a function of the susceptibility and of the volume of
the correlated domain. From here, we can infer the properties of SOC
upon the system, i.e., when driving it with $J\rightarrow 0^+$, the
system reacts by ''avalanches'' distributed in size according to a power
law
$$
P(s) \sim s^{-(1+\mu)}   \eqno{(4)}.
$$
The maximum avalanche size is related to $\xi$ by $s_{cutoff} \sim
\xi^D$ where $D$ is the fractal dimension of the avalanches. Then, the
flow caused by a local perturbation below $T_c$ is simply the average
size of avalanches, $\chi \sim \int_1^{s_{cutoff}} s P(s) ds \sim
s_{cutoff}^{1-\mu}$, yielding $\mu = 1 - \frac{\gamma}{\nu D}$. This
expression has been derived previously and checked with numerical
simulations [21] and yields $\mu \simeq 0.1$ in 2D and $\mu \simeq 0$ in
3D. In general, the avalanches are compact ($D=d$, where $d$ is the
space dimension), showing that $\mu$ can be determined from the
properties of the critical transition ($\gamma$ and $\nu$). The
determination of the dynamical exponent $z$, defined by the scaling of
an avalanche duration $t \sim \xi^z$, involves the exponent $\beta$ in
eq.(1). In the simplest picture where the dynamics is diffusive, the
sand flux is proportional to the diffusion coefficient which is itself
proportional to $\frac{\xi^2}{t}$. This expression assumes that there is
no renormalization of the microscopic transport coefficient entering in
the definition of the diffusion coefficient and leads to $z = 2 + \beta
\nu$. This last expression is uncertain in general since the simple
diffusion approximation is in question.

\subsubsection{Non-conservative sandpile models}
Our framework provides a simple and natural explanation for the
observation by [11] that
non-conservative sandpile models may be SOC. The authors present their
model as a toy
model for earthquakes. The local variable is the total force exerted
on a site. In their initial model, the force on each site is increasing very
slowly at a constant rate. When the force on a given site reaches a
threshold, it is
reset to zero while a fraction is redistributed on the nearest neighbors. The
non-conservation stems from the fact that the total amount which is
distributed is less
than the initial value.

Now, suppose that we work at fixed global force, i.e. we impose that the
sum over all
sites of the local forces is constant. Increasing this global force to a
larger value may
or may not trigger readjustments. The point again is that there exists a
critical value
for the global force above which the avalanches never stop, characterizing
a non-zero
average velocity $v$. Again, assuring $v\rightarrow 0^+$ places the system
at the
critical point by adjusting the global force to its critical value. It is
then clear why
the condition of conservation is not crucial  for the appearance of SOC :
SOC is seen to
rely fundamentally on the existence of an underlying sliding critical
point, which does
not need conservation, as well-documented in studies of the critical
dynamics of unstable
second order phase transitions [23].

\subsection{Earthquakes and tectonics}

It has been argued by several authors [24,25] that the earthquake
phenomenology in geology
is the signature of SOC. Let us thus consider a model elastic tectonic
plate [26,27],
scaled down in the laboratory to perform a mechanical thought experiment,
namely a shear
deformation imposed at its border for instance. The simplest situation is
where a
shear force $F$ is imposed on two opposite borders (the other two being
free), say, by a
spring set-up. (One could similarly consider a compressive experiment as
done in triaxial
tests [28], without any change in our discussion.) As the applied force $F$
increases, the
plate (which can contain pre-existing damage such as cracks and faults)
starts to deform
increasing the internal damage [29].
For sufficiently low $F$, after some transient during which the system
deforms and adjusts
itself to the applied force, the system becomes static and nothing happens
: the strain
rate or velocity of deformation becomes zero (here we are neglecting any
additional creep
or ductile behavior). As $F$ increases, the transient becomes longer and
longer since
larger and larger plastic-like deformations will develop within the plate.
There exists a
critical {\it plasticity} threshold $F_c$ at which the plate becomes globally
''plastic'', in the sense that it starts to flow with a non-zero strain rate
$\frac{d\epsilon}{dt}$ under fixed $F$. As $F$ increases above $F_c$, the
shear strain
rate $\frac{d\epsilon}{dt}$ increases. In models and in laboratory
experiments, this
plastic transition from a brittle to a ductile behavior is critical in the
usual sense.
$F$ is the control parameter and $\frac{d\epsilon}{dt}$ qualifies at the
order parameter
($\frac{d\epsilon}{dt}=0$ for $F<F_c$ and $\frac{d\epsilon}{dt}>0$ for
$F>F_c$).

Instead of controlling the force exerted on the system, let us apply
instead a constant
(very small) shear rate at the border of the plate. We thus recover the
''natural''
boundary condition for earthquakes and plate tectonic deformations (the
typical relative
plate velocity is of order $1$ cm/year compared to the fast rupture
velocity during an
earthquake of order $2$ km/sec.). Such a situation has been studied in many
works showing
the existence of SOC, both in the shape of power law earthquake size
distribution and in
the fractal fault geometry selected by the earthquake dynamics. It now
becomes clear that
this ''natural'' condition is nothing but driving the plate
 at  the critical point $F=F_c$ by controlling the order
parameter $\frac{d\epsilon}{dt}$ to a very small value, thus ensuring the
critical
properties of the systems. Note that SOC will appear if the corresponding
transition is critical. This may not be the case for some specific models
such as the spring-block models [30].

\subsection{Depinning transitions}

Consider an elastic line lying within a random system of pinning impurities
or asperities
and pulled by a force density per unit length or
at its two ends in a direction perpendicular to its average direction. This
can be a stretched or directed polymer with electric charges at its ends
which are
submitted to an electric field [31, 32], a vortex line in a
superconductor of type II with a superficial electric current which in the
presence of a
magnetic field will create a Lorentz force on the two ends of the vortex line
at
the sample borders [33], an interface created by a fluid
displacing another fluid in a porous medium [34] or a magnetic domain wall
in the
presence of quenched disorder.  The electric or magnetic field (the
field $E$ from now on) plays the role of the control parameter. For a given
(small) $E$
and some initial conditions, the overdamped dynamics will ensure the
relaxation of the
line in a well-defined line configuration. When increasing $E$
from some small value by some small increment, the line may stay fixed or
may readjust
itself via a sequence of localized sliding events into another static
conformation. As
$E$ increases, it has been well-documented that there is a critical value
$E_c$ above
which the line begins to move. The point $E=E_c$ is a {\it pinned-depinned}
critical
point very similar to usual dynamical critical points, endowed with all the
corresponding
properties. The order parameter is the average velocity $v$ of the line and
scales as $v
\sim (E-E_c)^{\beta}$.

Suppose now that instead of controlling $E$, one drives the line ends at a
constant
vanishingly small velocity, i.e. one controls the conjugate to the field,
namely the order parameter. By the same arguments as above, the line is then
automatically at its critical {\it pinned-depinned} transition point. As a
consequence,
long-range spatial correlations and large fluctuations appear, reflected in the
distribution of burst-like sliding events occurring along the line.

One can extract the ''avalanche'' distribution from numeric simulations on
an overdamped
elastic string in a random medium pulled by a force per unit length [31].
Using the fact
that the typical transverse fluctuation over a scale $z<\xi$ along the line
is of order
$z$ implies that the characteristic jump size for a portion of the line of
length $z$ is
proportional to $z$. Then, the average transverse motion over a correlation
length is
proportional to $\int_1^{\xi} z P(z) dz \sim \xi^{1-\mu}$, where $P(z) \sim
z^{-(1+\mu)}$
is the distribution of sliding events in the corresponding SOC problem. It
seems
reasonable to assume that this average transverse motion occurs over a time
scale
proportional to $\xi$ (it is easy to correct for any other scaling), which
leads to an
average velocity just above threshold scaling as $J \sim (E-E_c)^{\beta} \sim
\xi^{-\mu}$, giving the prediction $\mu = \frac{\beta}{\nu} \simeq 0.25$
using the
results $\beta \simeq 0.25$ and $\nu \simeq 1$ [31].

The same type of reasoning applies to a charged-density-wave (say an
elastic line pulled
in a direction parallel to its average direction) which is well-known to
exhibit a
{\em depinning} critical transition at some critical value of the driving
field [35].
Again, driving the CDW at a constant very small velocity creates a SOC
system with a
power law distribution of sliding events.

\subsection{Fractal growth processes}

It has been proposed that fractal growth processes, such as
diffusion-limited-aggregation (DLA), fluid imbibition as exemplified for
instance by
invasion percolation, dendritic growth, dielectric breakdown, rupture in
random media,
 constitute another class of systems exhibiting SOC [36-38].

These growth problems seem however to belong to a different class of
phenomena than
sandpile models since e.g. the internal geometrical structure of an aggregate
is
quenched (and only its perimeter is active) whereas the critical state of a
sandpile is
continuously rearranging under the action of external forcing.

Here, we want to point out that a stronger similarity emerges when using
the proposed
framework and the fractal self-organization of these growth processes can
be linked to the existence of an underlying critical point.
 We shall illustrate our ideas on the
annealed dielectric breakdown model in a cylindrical geometry (best suited
to define a
steady state regime), which is known to be equivalent to DLA [39]. The base of
the cylinder is fixed at potential $0$ and its other end is fixed at some
non-zero value
$V$. The rate of growth from the base is assumed to be a stochastic process
controlled
by a probability proportional to the electric field on the base growth
surface. The DLA
model is recovered in the limit where the growth is done quasi-statically.
This situation corresponds to fixing the growth rate (equivalently the
particle flux in
the DLA model) $J \rightarrow 0^+$.

Let us now consider the case where we impose a
constant average potential gradient $-\frac{dV}{dz}$, i.e. electric field
$E$, along the
cylinder axis. Furthermore, let us assume that dielectric breakdown, i.e.
growth on a
site, can only occur above a certain threshold, which is a quenched random
variable distributed according to a given distribution. This condition
ensures that the
growth problem now becomes similar to a pinned-depinned transition.
For those sites above their threshold, the rate of growth is again assumed
to be a
stochastic process controlled by a probability proportional to the electric
field.
If the applied electric field is very small, a few sites will break down
and the growth
cluster will evolve until all sites are below their threshold. This is the
{\em bound}
state. Increasing the applied electric field above a certain threshold
$E_c$ (a function
of the threshold distribution), the system begins to grow in an {\em
unbound} way at a
finite rate which increases as $E$ gets larger. The finite growth velocity
is determined
by the details of the dynamical breakdown processes (which we do not
describe here). Note
that a finite velocity corresponds to a finite particle flux in the DLA
model. For the critical transition, the control (resp. order) parameter is the
electric field or
particle concentration gradient (resp. the growth velocity or particle
flux). In the
regime above $E_c$, the growing clusters do not exhibit self-similarity at
all scales,
but a finite correlation length appears which is a decreasing function of
the growth
rate. In the language of fluid interfaces, pushed with an average velocity $c$,
the relevant dimensionless number is the ratio of the
Bernoulli velocity pressure $\rho v^2$ to the Laplace pressure
$\frac{\sigma}{b}$
 (similar to a Bond number), where
$\rho$ is the pushing fluid density, $\sigma$ is the surface tension and $b$
the
Hele-Shaw cell thickness or the interface radius of curvature.

In summary, the fractal structure of DLA clusters can be viewed
as the result of the quasi-static regime, i.e. the control of the order
parameter (growth
rate) of the corresponding critical transition at an infinitesimal
(positive)
value.

Similar mappings can be defined for the invasion percolation problem and
rupture
problems. In this respect, it is interesting to realize that the
quasi-static rupture
models extensively studied in the statistical literature [40] are
characterized by the
rupture of elements one by one, i.e. by controlling the rate of rupture (order
parameter) to a vanishingly small value. Truly dynamical models of rupture
[41] are in
this sense beyond the unstable critical point, separating the stable finite
damage phase
from the fully rupture phase.

\subsection{The Forest-Fire Model}

The self-organized critical forest-fire model introduced by Drossel \&
Schwabel [13,42] is another example of the proposed mechanism. The model is
defined on a d-dimensional lattice, where each site is either empty,
tree or fire. The lattice is updated synchronously according to the
following four rules: 1) A trees will grow on an empty site with
probability $p$, 2) Fire becomes empty 3) Fire spreads to n.n trees and
4) A tree catches fire spontaneously with probability $f$. The
existence of a critical point in the limit $f/p \rightarrow 0$ is
expected, since the average number of trees destroyed by a lightning is
$\left< s \right> = (f/p)^{-1}(1-\rho_t)/\rho_t$ ($\rho_t$ is the mean
tree density) provided $f \ll p \ll 1/T(s)$ [42], where $T(s)$ is the
average time it takes to burn a tree cluster of size $s$. This
separation of time scales is again nothing but a condition of {\em slow
driving} common for SOC-models. Furthermore, $\rho_t < 1$ in order for
$\left< s \right>$ to diverge. This then suggests $\rho_t$ as the control
parameter for the critical transition in agreement with the proposed
framework. Changing the first rule of the model allows one to tune
$\rho_t$ to its critical value: Each time a tree is burned down a new
tree is put at random thus keeping $\rho_t$ fixed to a controlled value.
The order parameter is then the density of fires $\rho_f$, the condition
$f/p \rightarrow 0$ above corresponding to $\rho_f = 0^+$. In the case
of $f=0$, the forest-fire can only propagate on dynamically
connected tree clusters, thus defining a critical tree density below
which forest-fire will finally extinguish itself and burn forever above.
It is then clear that the parameter $f$ plays the role of an
external field thus modifying the transition and explaining the
systematic deviations from a pure power law seen in simulations [42,43].

\section{Concluding remarks}

\begin{enumerate}

\item We have proposed a general conceptual framework for self-organized
criticality,
which consists in mapping SOC onto unstable critical points, controlled
by driving the corresponding order parameter at an infinitesimal value.

Our present theory clarifies and extends the previously
recognized mechanism of a feedback of the order parameter on the control
parameter as
being essential for SOC [19].

The mapping between SOC and usual critical transitions offers a novel but
natural
route to study further the properties associated to SOC, namely by
characterizing the
critical properties of the underlying critical point itself. In a way, we
have only
displaced the search for the underlying mechanism for SOC to that of the
appearance of {\em unstable} critical points. However, we feel that this
is a significant
improvement for two reasons : 1) some of the underlying unstable critical
points are
known and have been documented per se. Their knowledge can thus bring new
light on the
SOC models. In particular, the present framework illuminates the physical
meaning of the
slow driving common to all systems exhibiting SOC. 2) Even if the
underlying critical
point is new, its study can probably be quite efficient by employing the
large toolbox
developed in the last twenty or more years in this field.

\item {\em Supercritical bifurcations :} Note that our framework applies
directly to
supercritical and Hopf bifurcations, when considering the situation where
the order
parameter is controlled at the value $0^+$. This situation can be modelled
analytically
by writing a general (possibly complex) Landau-Ginzburg equation for the
order parameter
fluctuations, conditioned to have a vanishingly small average order
parameter. This is
left for the future.

\item {\em Renormalization group and fixed-scale transformation :}
Controlling the order
parameter $J \rightarrow 0^+$ of an unstable critical point
does not
allow for a standard renormalization group procedure. Indeed, in this
situation, there is
no control parameter and the critical exponents cannot be obtained by the
standard
procedure in terms of the derivatives of the renormalization group
transformation of the
control parameters under scale transformation. Thus, the exponents related
to the
distance from the critical point do not exist. Let us stress that this is
not due to a
special attractive property of the critical point, as claimed in [44], but
results solely
from the special driving conditions which ensures the exact positioning of
the dynamics
on the unstable critical point. In this situation, a generalized
renormalization
procedure should reflect the exact positioning on a critical point, i.e.
should give
the signature of an {\em attractive} critical point. This is indeed born
out in the fixed
scale transformation procedure introduced by [44]. We believe that finding a
general renormalization theory for systems right at their critical point
could provide a
new class of tools for general critical phenomena. This could be related to
conformal invariance theory [45].

\item {\em Relation with Goldstone modes :} Our approach allows us to
clarify the proposal
[46] that SOC stems from the non-linear dynamics of Goldstone modes,
however by returning
the logic of Obukhov's argument : criticality in SOC is {\em not} the effect of
interaction of Goldstone gapless modes as claimed; the gapless modes result
rather from
the underlying unstable critical point, stabilized by the special driving
condition.
Let us recall that, in the long-wavelength limit, Goldstone modes
reduce to a homogeneous  displacement of the sample or to a uniform
rotation of the
whole spin system. Since a SOC system is right at the
critical point of a standard unstable critical phase transition
characterized by a
spontaneous symmetry breaking, the avalanches can be viewed as nothing but
the Goldstone
fluctuations (i.e. large scale displacements) attempting to restore the
broken symmetry.
In the case of a discrete symmetry breaking, the avalanches correspond to
droplet fluctuations [47].

\item {\em Relation with singular diffusion :} The singular diffusion
property of
continuous equations obtained by taking the hydrodynamic limit of SOC
models [48] derives
straightforwardly from our framework, since it is the direct signature of
the precise
localization at an unstable critical point. Thus, all
theories of SOC in
terms of singular diffusion [48,49] are
only the expression that the governing equation is that of a system sitting
precisely at
a critical point. Let us recall that even more generally, singular
diffusion occurs on the
approach to any supercritical bifurcation, the best-known example being the
Rayleigh-B\'enard instability. In this case, the order parameter is the average
convection velocity and the fluctuations are associated to streaks or
patches of non-zero
velocity  occurring below the critical Rayleigh number $R$ at which global
convection
starts off. The spatial diffusion coefficient $D(R)$ diverges as $D(R) \sim
(R_c -
R)^{-\frac{3}{2}}$ and reflects the existence of very large velocity
fluctuations. A
scaling argument can be written to get this powerlaw [50]. Similarly, the
singular diffusion found in SOC models can also be derived from similar
scaling reasoning
[51], showing the common origin of the singular diffusion.

\item Our proposed framework is not in contradiction with the recent
recognition by
several authors [8,15-18] that SOC is deeply connected to the
mechanism of partial synchronization of relaxation oscillators with
thresholds. Under a
slow driving of the order parameter, each threshold element (for instance
a column in the sandpile) taken in isolation undergoes periodic oscillations of
relaxation. The complete problem corresponds to describe the result of the
coupling
between these oscillations, in terms of a competition between
synchronization and
desynchronization effects. In other words, the dynamics of relaxation
oscillators
describes the detailed response of the critical point under the slow
driving of the order
parameter.

\end{enumerate}

\pagebreak

{\Large \bf References}

[1] V. Pareto, Cours d'\'economie politique. Reprinted as a volume of
{\it Oeuvres Compl\`etes} (Droz, Geneva, 1896-1965).

[2] B. V. Gnedenko, A.N. Kolmogorov, {\it Limit distributions for sum
of independent random variables}, Addison Wesley, Reading, MA, (1954).

[3] P. L\'evy, Th\'eorie de l'addition des variables al\'eatoires
(Gauthier Villars, Paris,1937-1954).

[4] E. Montroll, M. Shlesinger, `On the Wonderful world of Random Walks', in
Nonequilibrium phenomena II, From stochastic to hydrodynamics, Studies in
statistical
mechanics XI (J.L. Lebowitz and E.W. Montroll, eds) (North Holland,
Amsterdam, 1984).

[5] B.B. Mandelbrot, `The Fractal Geometry of Nature' (Freeman, San
Francisco, 1983).

[6] E. Fama, Management Science 11, 404 (1965).

[7] P. Bak, C. Tang, K. Wiesenfeld, Phys. Rev. Lett. 59, 381 (1987);
 Phys. Rev. A 38, 364 (1988).

[8] D. Sornette, Les ph\'enom\`enes critiques auto-organis\'es,
Images de la Physique 1993, \'edition du CNRS.

[9] Hwa T. and Kardar M., Phys.Rev.Lett. 62 , 1813 (1989)

[10] Grinstein G., Lee D.-H and Sachdev S., Phys.Rev.Lett. 64 , 1927 (1990);
        Grinstein G. et Lee D.-L., Phys.Rev.Lett. 66, 177 (1991)

[11] Z. Olami, H.J.S. Feder and K. Christensen,  Phys.Rev.Lett. 68, 177 (1991);
K. Christensen and Z. Olami,  Phys.Rev.A46, 1829 (1992)

[12] P. Bak, K. Chen and M. Creutz, Nature (London) 342, 780 (1989);
P. Alstrom and J. Leao, Phys. Rev. E49, R2507 (1994)

[13] B. Drossel and F. Schwabl,  Phys.Rev.Lett. 69, 1629 (1992)

[14] Bak P. and K. Sneppen, Phys. Rev.Lett. 71, 4083 (1993);
Flyvbjerg H., K. Sneppen and P. Bak, Phys. Rev.Lett. 71, 4087 (1993);
Ray T.S. and N. Jan, Phys. Rev.Lett. 72, 4045 (1994)

[15] K. Christensen, Self-organization in models of sandpiles, earthquakes and
flashing fireflies, PhD Thesis,  (Nov. 1992)

[16] A. Corral, C.J. P\'erez, A. D\'iaz-Guilera and A. Arenas, Self-organized
criticality and synchronization in a lattice model of integrate-and-fire
oscillators,
preprint (1994)

[17] A.A. Middleton and C. Tang, Self-organized criticality in
non-conserved systems,
preprint (1994)

[18] S. Bottani, Pulse-coupled relaxation oscillators : from biological
synchronization
to self-organized criticality, preprint (1994)

[19] D.Sornette, J.Phys. I France 2, 2065 (1992);
        N.Fraysse, A.Sornette and D.Sornette, J.Phys. I France 3, 1377 (1993)

[20] J. Rajchenbach,  Phys. Rev.Lett. 65, 2221 (1990)

[21] C. Tang and P. Bak, Phys.Rev.Lett. 60, 2347 (1988)

[22] Dhar D. et Ramaswamy R., Phys.Rev.Lett. 63, 1659 (1989);
Dhar D., Phys.Rev.Lett. 64, 1613 (1990)

[23] Hohenberg P.C. and Halperin B.I.,  Rev.Mod.Phys. 49, 435 (1977)

[24] Sornette, A. and D. Sornette, Europhys. Lett. 9, 197 (1989);
Sornette D., Self-organized criticality in plate tectonics, in Proceeding
of the NATO ASI,
Spontaneous formation of space-time structures and criticality, eds. by
Riste T. and
Sherrington D., Geilo, Norway 2-12 April 1991 (Kluwer Academic Press,
1991), p.57.

[25] Bak P. and Tang C., J. Geophys. Res. 94, 15635 (1989);
K. Ito and M. Matsuzaki, J. Geophys. Res. 95, 6853 (1990)

[26] Cowie P., C.Vanneste and D.Sornette, J.Geophys.Res. 98, 21809 (1993);
Miltenberger P., D. Sornette and C. Vanneste, Phys.Rev.Lett. 71, 3604
(1993);
Sornette D., P. Miltenberger and C.Vanneste, Pageoph 142, 491 (1994).

[27] A. Cochard and R. Madariaga, Pageoph 142, 419 (1994)

[28] Evesque P., J.Phys.France 51, 2515 (1990);
D. Sornette, A. Sornette and P. Evesque, Frustration and disorder in
granular media and
tectonic blocks : implications for earthquake complexity, Nonlinear Processes
in
Geophysics, in press (1994)

[29] Lockner D.A., J.D. Byerlee, V. Kuksenko, A. Ponomarev and A. Sidorin,
Nature 350, 39 (1991)

[30] Burridge R. and Knopoff L., Bull. Seismol. Soc. Am. 57, 341 (1967);
Carlson J.M. and Langer J.S., Phys.Rev.Lett. 62, 2632 (1989);
Schmittbuhl J., Vilotte J.-P. and Roux S., Europhys.Lett. 21, 375 (1993)

[31] M. Dong, M.C. Marchetti, A.A. Middleton and V. Vinokur,
Phys.Rev.Lett. 70, 662 (1993)

[32] M. M\'ezard, J.Phys.France 51, 1831 (1990)

[33] C. Tang, S. Feng and L. Golubovic, Phys.Rev.Lett. 72, 1264 (1994)

[34] Horv\'ath V.K., F. Family and T. Vicsek, J.Phys. A24, L25 (1991)

[35] D.S. Fisher, Phys. Rev. B31, 1396 (1985)

[36] P. Alstrom, Phys.Rev. A 38,4905 (1988); Phys.Rev. A 41,
7049 (1990)

[37] Bak P. and K. Chen, Physica D38, 5 (1989)

[38] I.M. J\'anosi and A. Czir\'ok, Fractals 2, 153 (1994)

[39] L. Niemeyer, L. Pietronero and H.J. Wiesmann, Phys.Rev.Lett. 22, 1033
(1984);
J. Kert\'esz, Dielectric breakdown and single crack models, in ref.[40], p. 261

[40] H.J. Herrmann and Roux S., eds., Statistical models for the
fracture of disordered media (Elsevier, Amsterdam, 1990)

[41] D. Sornette and Vanneste C., Phys.Rev.Lett. 68, 612 (1992);
D. Sornette, C. Vanneste and L. Knopoff, Phys.Rev.A 45, 8351 (1992);
C. Vanneste and D. Sornette, J.Phys.I France 2, 1621 (1992)

[42] S. Clar, B. Drossel \& F. Schwabel, Phys.Rev.E 50, 1009 (1994)

[43] K. Christensen, H. Flyvbjerg \& Z. Olami, Phys.Rev.Lett. 71,
2737 (1993)

[44] L. Pietronero, A. Vespignani and S. Zapperi, Phys.Rev.Lett. 72, 1690
(1994)

[45] J.L. Cardy, Physica A140, 219 (1986)

[46] S. Obukhov, Phys.Rev.Lett. 65, 1395 (1990)

[47] Huse D. and Fisher D.S., Phys.Rev.B35, 6841 (1987)

[48] Carlson J.M., Chayes J.T., Grannan E.R. an Swindle G.H.,
Phys.Rev.Lett. 65, 2547
(1990)

[49] B\'antay P. and I.M. J\'anosi, Phys. Rev. Lett. 68, 2058 (1992)

[50] D. Sornette, J.Phys.I France 4, 209 (1994)

[51] Kadanoff L.P., A.B. Chhabra, A.J. Kolan, M.J. Feigenbaum, I.
Procaccia, Phys. Rev.
A45, 6095 (1992).

\end{document}